\documentclass{elsart1p}

\usepackage{amssymb,amsmath}
\usepackage{wrapfig}
\usepackage{graphicx,graphics}

\def\Dsl{{\rlap{\kern2.25pt /}{D}}}
\def\Asl{{\rlap{\kern2.25pt /}{A}}}
\def\dsl{{\rlap{\kern0.5pt /}{\partial}}}
\def\xisl{{\rlap{\kern0.5pt /}{\xi}}}
\def\asl{{\rlap{\kern0.5pt /}{a}}}
\def\bsl{{\rlap{\kern0.5pt /}{b}}}

\def\Tr{{\rm Tr}}
\def\({\left(}
\def\){\right)}

\begin{document}

\begin{frontmatter}



\title{Exotic phases of finite temperature $SU(N)$ gauge theories}


\author[label1]{Joyce C. Myers} and \author[label2]{Michael C. Ogilvie}

\address[label1]{Department of Physics, Swansea University, Singleton Park, Swansea SA2 8PP, UK}
\address[label2]{Department of Physics, Washington University, One Brookings Dr., Campus Box 1105, St. Louis, MO, 63130, USA}

\begin{abstract}
We calculate the phase diagrams at high temperature of SU(N) gauge theories with massive fermions by minimizing the one-loop effective potential. Considering fermions in the adjoint (Adj) representation at various $N$ we observe a variety of phases when $N_f \ge 2$ Majorana flavours and periodic boundary conditions are applied to fermions. Also the confined phase is perturbatively accessible. For $N = 3$, we add Fundamental (F) representation fermions with antiperiodic boundary conditions to adjoint QCD to show how the $Z(3)$-symmetry breaks in the confined phase.
\end{abstract}

\begin{keyword}
finite temperature QCD
\PACS 11.10.Wx \sep 12.38.Bx \sep 12.38.Gc
\end{keyword}
\end{frontmatter}

\begin{section}{Introduction}

The phase diagram of pure $SU(N)$ gauge theories is defined by a confined phase in which the $Z(N)$ center symmetry is preserved, and a deconfined phase in which the center symmetry is spontaneously broken. Lattice simulations have been particularly important in the study of the confined phase because it is in a region of strong coupling where perturbation theory is not valid. In this paper we discuss theories which are QCD-like, given by two $Z(N)$-invariant extensions to pure Yang-Mills theory: 1) center-stabilized Yang-Mills theory, which introduces multiply wound adjoint Polyakov loops to Yang-Mills theory, and 2) adjoint QCD [QCD(Adj)], which is QCD with adjoint representation fermions rather than fundamental.

In \cite{Myers:2007vc} we performed lattice simulations with an adjoint Polyakov loop extension to Wilson action. The action has the form:

\begin{equation}
S = S_W + \sum_{{\vec x}} H_A \Tr_A P({\vec x}) ; \hspace{1cm} S_W (\beta, U) = - \beta \sum_p \( 1 - \frac{1}{N} {\rm Re} \Tr_F U_p \) ,
\label{lattice_action}
\end{equation}

\begin{wrapfigure}{r}{0.38\textwidth}
  \begin{center}
    \includegraphics[width=5.2cm]{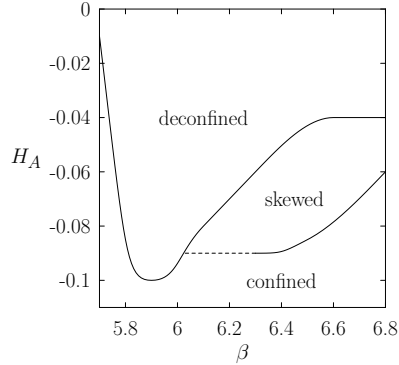}
  \end{center}
  \caption{Phase diagram in $SU(3)$ for lattice simulations of YM theory extended with a $Tr_A P$ term. The pure gauge theory result corresponds to $H_A = 0$.}
  \label{phase_diag}
\end{wrapfigure}

\noindent where $\sum_{{\vec x}}$ is over all spatial sites. Figure \ref{phase_diag} shows the resulting phase diagram for $SU(3)$ as a function of the inverse coupling $\beta = 2 N / g^2$ and the tunable parameter $H_A$. Two important features of this phase diagram are that the confined phase is accessible in a region of weak coupling, and that there is a region of the parameter space in which a new, "skewed" phase is favoured over both the confined and deconfined phases. Both of these features were recently confirmed in \cite{Wozar:2008nv} in simulations using a demon algorithm. Simulations of a similar theory in \cite{Dumitru:2007ir} also confirm that the confined phase becomes accessible beyond the deconfinement temperature of the pure gauge theory.

\end{section}
\begin{section}{Center-stabilized Yang-Mills theory}

The simulation results using eq. (\ref{lattice_action}) are in good agreement with high-temperature perturbation theory \cite{Myers:2007vc}. However, for $SU(4)$ both lattice results and perturbation theory agree that the confined phase is not observed in the weak-coupling limit, unlike in $SU(3)$. In order to perturbatively obtain the confined phase for arbitrary $N$ we introduced in \cite{Ogilvie:2007tj} an extension in terms of the Polyakov loop $P = {\rm diag} \{ e^{i v_1}, e^{i v_2}, ..., e^{i v_N} \}$ to the boson contribution \cite{Gross:1980br} from pure Yang-Mills theory

\begin{equation}
V_{CYM} (P) = - \frac{2}{\pi^2 \beta^4} \sum_{n=1}^{\infty} \frac{1}{n^4} \left[ \Tr_A \( P^n \) \right] + \frac{1}{\beta} \sum_{n=1}^{N / 2} a_n \Tr_F \( P^n \) \Tr_F \( P^{\dagger n} \) ,
\label{cym}
\end{equation}

\noindent where $\lfloor N / 2 \rfloor$ is the integer part of $N / 2$. This is the minimum number of terms required to obtain the confined phase for some value of the $a_n$ parameters. This potential was recently extensively studied in \cite{Unsal:2008ch} and we have adopted their notation and the nomenclature "center-stabilized Yang-Mills theory". We minimized $V_{CYM}$ with respect to the eigenvalue angles $v_i$ to obtain the phase diagram for a range of values of the $a_n$.

\end{section}
\begin{section}{Adjoint QCD}

In order to have a renormalizable theory we also study adjoint QCD where periodic boundary conditions (PBC) are applied to fermions in the adjoint representation. The lattice action in eq. ( \ref{lattice_action} ) and the center-stabilized theory of eq. ( \ref{cym} ) are both approximations to adjoint QCD with PBC. The one-loop effective potential for $N_f$ Majorana flavours ($N_{f, Dirac} = \frac{1}{2} N_f$) of fermions in representation $R$ and with finite mass $m$ is \cite{Meisinger:2001fi}

\begin{equation}
\begin{aligned}
V_{1-loop} &= \frac{1}{\beta V_3} \left[ - N_f \ln \det \( - D_R^2 (P) + m^2 \) + \ln \det \( - D_{adj}^2 (P) \) \right]\\
&= \frac{m^2 N_f}{\pi^2 \beta^2} \sum_{n=1}^{\infty} \frac{( \pm 1 )^n}{n^2} {\rm Re} \left[ \Tr_R \( P^n \) \right] K_2 \( n \beta m \) - \frac{2}{\pi^2 \beta^4} \sum_{n=1}^{\infty} \frac{1}{n^4} \Tr_A \( P^n \) .
\end{aligned}
\label{adjQCD}
\end{equation}

\begin{figure}
  \hfill
  \begin{minipage}[t]{.45\textwidth}
    \begin{center}  
      \includegraphics[width=0.95\textwidth]{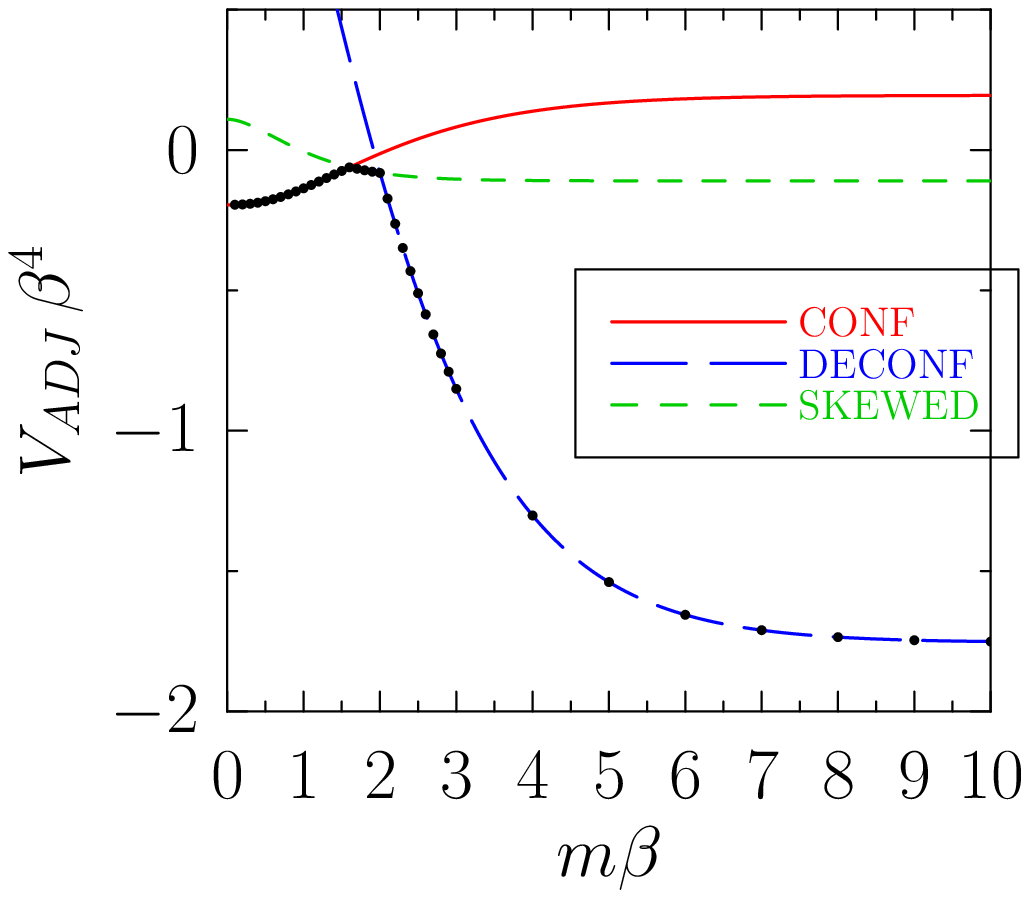}
    \end{center}
  \end{minipage}
  \hfill
  \begin{minipage}[t]{.45\textwidth}
    \begin{center}
\includegraphics[width=0.94\textwidth]{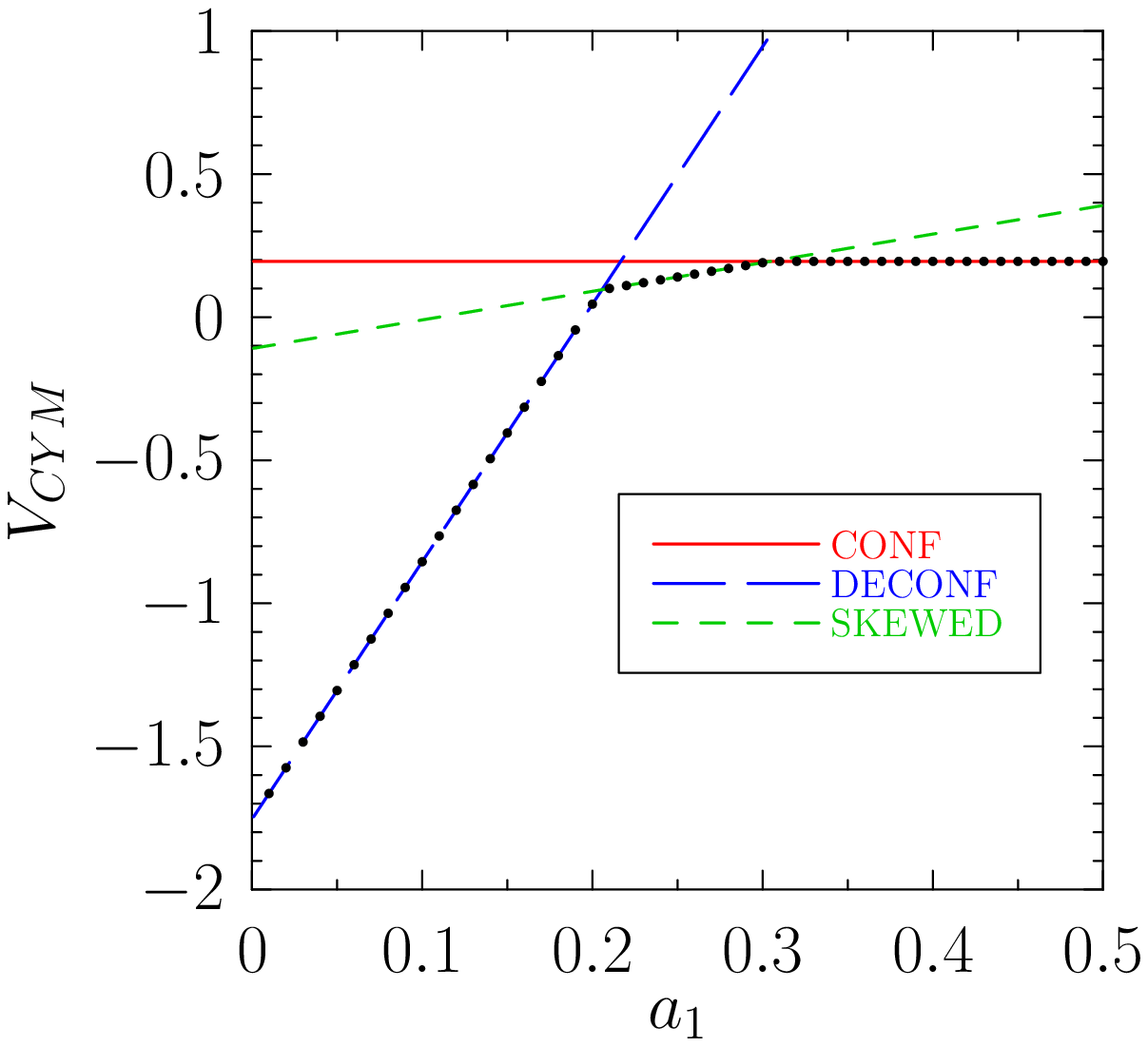}
    \end{center}
  \end{minipage}
  \hfill
  \caption{$N = 3$: (Left) $V_{ADJ(+)}$, $N_f = 2$ Majorana flavours; (Right) $V_{CYM}$ vs. $a_1$}
  \label{adj_n3}
\end{figure}

We minimize $V_{1-loop}$ with respect to the Polyakov loop eigenvalue angles $v_i$ to obtain the phase diagram as a function of $m \beta$. The resulting phase diagram of adjoint QCD is quite rich in the case of $N_f \ge 2$ Majorana flavours ($N_{f, Dirac} \ge 1$) and PBC on fermions \cite{Myers:2008ey}. The phase diagram for $N = 3$ and $N_f = 2$ is shown in Figure \ref{adj_n3} (L). The dots correspond to the results of the numerical minimization of $V_{1-loop}$ in eq. ( \ref{adjQCD} ). Similarly, the phase diagram of the center-stabilized theory is shown in Figure \ref{adj_n3} (R). The phase curves in Figure \ref{adj_n3} can be classified according to the Polyakov loop eigenvalue angles: ${\bf v} = \{ v_1, v_2, ..., v_N \}$. For $N = 3$ the confined phase is defined by ${\bf v} = \{ 0, \frac{2 \pi}{3}, \frac{4 \pi}{3} \}$. The deconfined phases are defined by ${\bf v} = \{ 0, 0, 0 \}$ and $Z(3)$ rotations. The skewed phases are $SU(2) \times U(1)$ phases defined by ${\bf v} = \{ 0, \pi, \pi \}$ and $Z(3)$ rotations.

\end{section}
\begin{section}{QCD with fermions in the adjoint and fundamental representations}

The perturbative accessibility of the confined phase in both the center-stabilized theory and in adjoint QCD for $N_f \ge 2$ Majorana flavours is useful for studying how the $Z(3)$ symmetry is broken when adding fundamental fermions with antiperiodic boundary conditions (ABC). We opt to add the fundamental fermions to adjoint QCD to have a renormalizable theory. For $N_{F} = 3$ Dirac flavours of fundamental fermions asymptotic freedom is maintained for up to $N_A = 4$ Majorana flavours of adjoint fermions.


\begin{figure}
  \hfill
  \begin{minipage}[t]{.45\textwidth}
    \begin{center}  
      \includegraphics[width=0.86\textwidth]{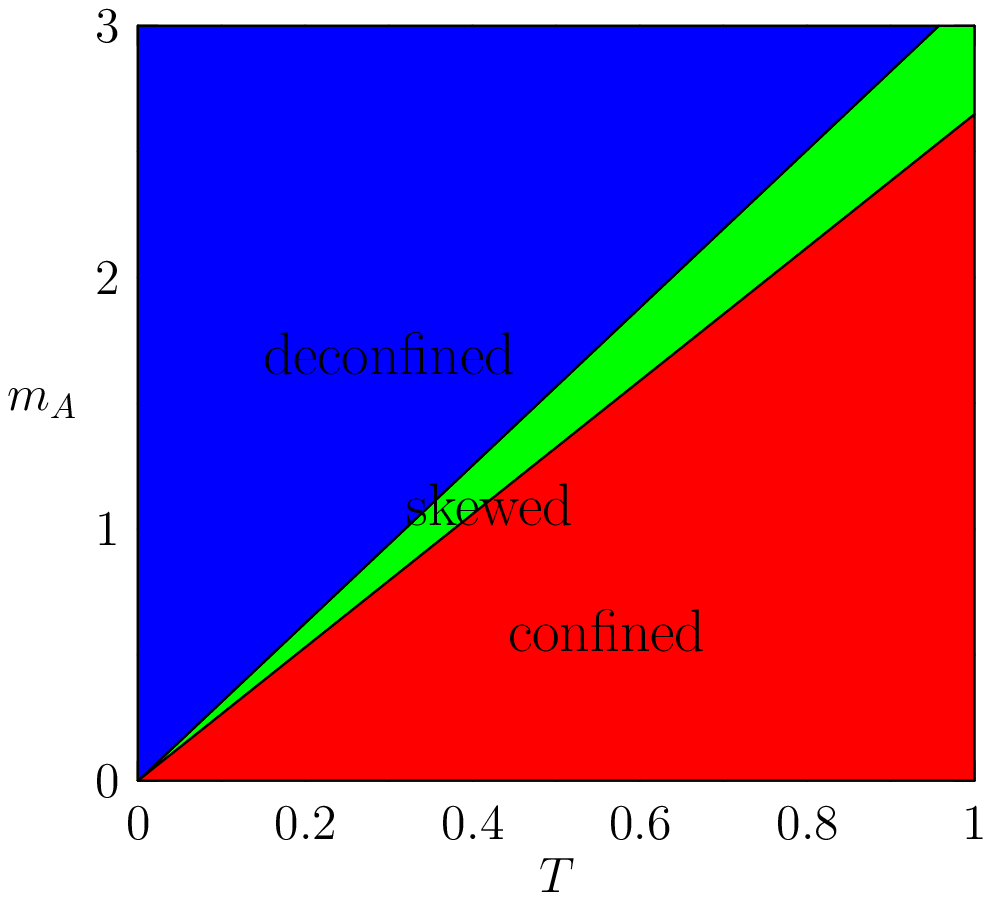}
    \end{center}
  \end{minipage}
  \hfill
  \begin{minipage}[t]{.45\textwidth}
    \begin{center}
\includegraphics[width=0.9\textwidth]{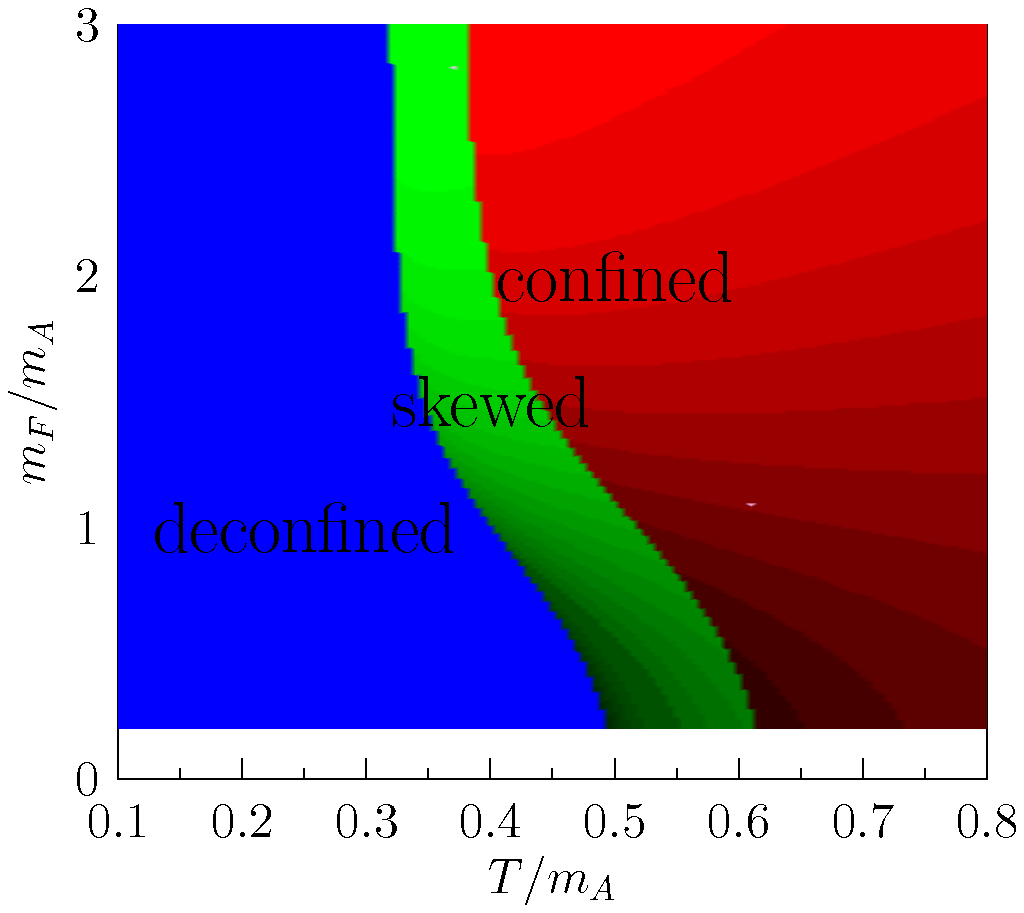}
    \end{center}
  \end{minipage}
  \hfill
  \caption{$N = 3$: (Left) Phase diagram of adjoint QCD with $N_A = 4$ Majorana flavours. (Right) The phase diagram of $V_{QCD(F_{(-)} A_{(+)})}$ at high temperature as it varies with $m_F$ ($N_F = 3$, $N_A = 4$, $m_A = 1$).}
  \label{qcd_FA}
\end{figure}

Figure \ref{qcd_FA} (L) shows the phase diagram for $N = 3$ from minimizing eq. (\ref{adjQCD}) for $N_A = 4$ Majorana flavours of adjoint fermions and PBC. Figure \ref{qcd_FA} (R) shows effect of adding $N_F = 3$ Dirac flavours of fundamental fermions with antiperiodic boundary conditions to adjoint QCD. Here the mass of the adjoint fermions is fixed at $m_A = 1$. As the mass of fundamental fermions $m_F$ is brought down from infinity the eigenvalues of the confined phase shift so that the $Z(3)$ symmetry is broken. $m_F = \infty$ corresponds to adjoint QCD and the confined phase is given by ${\bf v} = \{ 0, - \phi, \phi \}$ where $\phi = 2 \pi / 3$ and $\Tr_F P = 0$. The contours in the confined phase of Figure \ref{qcd_FA} (R) represent intervals of $\Delta \phi = \pi / 36$ away from ${\bf v} = \{ 0, - 2\pi / 3, 2 \pi / 3 \}$. As $m_F \rightarrow 0$, $\phi$ decreases towards roughly $\pi / 3$, depending on $T$, and $\Tr_F P$ goes out from $0$ along the real axis towards 2, keeping the confined phase distinguishable at any observed $m_F$ from the deconfined phase which has $\Tr_F P = 3$.

\end{section}
\begin{section}{Conclusions}

We study three $Z(N)$-invariant Polyakov loop extensions to Yang-Mills theory that offer confinement in a perturbatively accessible regime, as well as additional phases under certain conditions. An adjoint Polyakov loop extension gives perturbative confinement for $N = 3$. Center-stabilized Yang-Mills theory includes the minimum number of powers of adjoint Polyakov loop terms needed to get perturbative confinement for all $N$. As well this theory contains various other phases depending on the values of the $a_n$ parameters. Adjoint QCD for $N_A  = 2$ or more Majorana flavours and PBC on fermions also gives perturbative confinement for all $N$, and small $m \beta$, in addition to other phases contained in the center-stabilized model. Adding fundamental representation fermions to adjoint QCD, while being careful to preserve asymptotic freedom, shows that in the confined phase the $Z(3)$ symmetry is broken in a predictable way. The degree of symmetry breaking depends on the mass of fundamental fermions $m_F$ and the temperature, but the confined phase remains distinguishable from the deconfined phase for all observed $m_F \beta$.

\end{section}



\end{document}